\documentclass{article}
\usepackage{spconf,amsmath,graphicx}
\usepackage{booktabs}

\title{LightSpeech: Lightweight and Fast Text to Speech with \\ Neural Architecture Search}
\name{Renqian Luo$^1$, Xu Tan$^2$, Rui Wang$^2$, Tao Qin$^2$, Jinzhu Li$^3$, Sheng Zhao$^3$, Enhong Chen$^1$ and Tie-Yan Liu$^2$}
\address{$^1$University of Science and Technology of China, $^2$Microsoft Research Asia, $^3$Microsoft Azure Speech\\
$^1$lrq@mail.ustc.edu.cn,cheneh@ustc.edu.cn,\\$^2$\{xuta,ruiwa,taoqin,tyliu\}@microsoft.com,$^3$\{jinzl,sheng.zhao\}@microsoft.com}
\begin{document}
%
\maketitle
\begin{abstract}
Text to speech (TTS) has been broadly used to synthesize natural and intelligible speech in different scenarios. Deploying TTS in various end devices such as mobile phones or embedded devices requires extremely small memory usage and inference latency. While non-autoregressive TTS models such as FastSpeech have achieved significantly faster inference speed than autoregressive models, their model size and inference latency are still large for the deployment in resource constrained devices. In this paper, we propose LightSpeech, which leverages neural architecture search~(NAS) to automatically design more lightweight and efficient models based on FastSpeech. We first profile the components of current FastSpeech model and carefully design a novel search space containing various lightweight and potentially effective architectures. Then NAS is utilized to automatically discover well performing architectures within the search space. Experiments show that the model discovered by our method achieves 15x model compression ratio and 6.5x inference speedup on CPU with on par voice quality. Audio demos are provided at https://speechresearch.github.io/lightspeech.
\end{abstract}
\section{Introduction}
\label{sec:intro}
Text to speech (TTS) has been widely used to synthesize natural and intelligible speech audio given text, and has been deployed in many services such as audio navigation, newscasting, tourism interpretation, etc. While neural network based TTS models~\cite{tacotron,tacotron2,deepvoice3,transformertts} have greatly improved the voice quality over conventional TTS systems, they usually adopt autoregressive generation with large inference latency, which make them difficult to deploy on various end devices such as mobile phones or embedded devices. Recently, non-autoregressive TTS models~\cite{fastspeech,peng2019parallel,lim2020jdi,flowtts,fastspeech2} have significantly accelerated the inference speed over previous autoregressive systems. Despite their success, the models still have relatively large model size, inference latency and power consumption when deploying in resource constrained scenarios~(e.g., mobile phones, embedded devices and low budget services where CPU is mainly available).

There have been many techniques on designing lightweight and efficient neural networks, such as shrinking, tensor decomposition~\cite{tensordecomposition}, quantization~\cite{binaryconnect} and pruning~\cite{channelpruning}. These methods have achieved significant success on compressing big models into smaller ones with less computational cost. However, most of them are designed for convolution neural networks for computer vision tasks, which involve many area specific knowledge or characteristics, and cannot be easily extended to sequence learning tasks~(e.g., natural language processing and speech processing) with recurrent neural networks, attention networks, etc. For example, manually reducing the depth and width of the network brings severe performance drop~(as in Table~\ref{tbl:cmos}). Recently, neural architecture search~(NAS)~\cite{nasnet,nao} is leveraged to automatically design light weight models with promising performance~\cite{efficientnet,adabert}. However, applying NAS to a new area and task is challenging which requires careful design and chosen of the search space, the search algorithm and the evaluation metric: 1) The search space determines the potential bound of the performance and requires carefully design. A good search space is efficient to search and a poorly designed search space is hard to find promising architectures. 2) The search algorithm should be carefully chosen to fit and adapt to the task with specific characteristics. 3) The metric should also be designed or chosen to best represent the final evaluation metric. Directly applying existing NAS algorithms may lead to no improvements.

In this work, we propose LightSpeech, which leverages NAS for designing lightweight and fast TTS models with much smaller size and faster inference speed on CPU. Firstly, we carefully profile the bottlenecks of current TTS model~\cite{fastspeech2}. Secondly, according to the observations from the profiling, we design a novel search space that covers a range of lightweight models. Thirdly, among many well-performing NAS algorithms, we adopt accuracy prediction based NAS~\cite{neuralpredictor,gbdtnas} which is straightforward and efficient. Specifically, we adopt GBDT-NAS~\cite{gbdtnas} to perform the search due to its promising performance and fitness in our task~(the chain-structure of our search space is well fitted by GBDT-NAS).

Experiments show that compared to the original FastSpeech 2 model~\cite{fastspeech2}, architecture discovered by our LightSpeech achieves 15x compression ratio~(1.8M vs. 27M), 16x less MACs~(0.76G vs. 12.50G) and 6.5x inference time speedup on CPU.

\section{Method}
\label{sec:method}
Considering FastSpeech~\cite{fastspeech,fastspeech2} is one of the most popular non-autoregressive TTS models with fast and high-quality speech synthesis, we mainly adopt it as the model backbone. First we analyze the memory and latency of each component in the current models in Section~\ref{sec:analysis}. Then we design a novel search space including various operations in Section~\ref{sec:search_space}. Finally we introduce the search algorithm used to find efficient models in Section~\ref{sec:search_algorithm}.

\subsection{Profiling the Model}
\label{sec:analysis}
Non-autoregressive TTS models~\cite{fastspeech, fastspeech2,lim2020jdi,flowtts,peng2019parallel} generate speech in parallel and greatly speed up the inference process compared to autoregressive TTS models~\cite{tacotron,tacotron2,deepvoice3}. However, the models are still large which cause high memory usage and inference latency when deploying on end devices with limited computing resources~(e.g., mobile phones and embedded devices). For example, taking FastSpeech 2~\cite{fastspeech2} (which further improves the voice quality of FastSpeech by introducing more variance information) as an example, it has 27M parameters with more than 100M memory footprint and is 10x slower on CPU than on GPU\footnote{It takes $5.6\times 10^{-3}$ seconds and $6.1\times 10^{-2}$ seconds to generate one second waveform on GPU and CPU respectively.}.

In order to determine the network backbone and the search space, we profile each component in FastSpeech 2 model to identify the bottlenecks of memory and inference speed. The core model of FastSpeech 2 contains 5 parts: encoder, decoder, duration predictor, pitch predictor and energy predictor. The encoder and the decoder consists of 4 feed-forward Transformer blocks~\cite{fastspeech} respectively. The duration predictor is a 2-layer 1D-convolution neural network with kernel size $3$. The pitch predictor is a 5-layer 1D-convolution neural network with kernel size $5$. The energy predictor has the same structure as the pitch predictor. We measure the size~(i.e., number of parameters) and the inference speed of each component as shown in Table~\ref{tbl:analysis}.

\begin{table}[htbp]
\small
\centering
\begin{tabular}{l|lrc}
\toprule
Name          & Structure  & \#Params & RTF  \\
\midrule
FastSpeech 2     & -  & 27.02M & $6.1\times 10^{-2}$ \\
\midrule
Encoder       & 4 Trans Block & 11.56M & $8.1\times 10^{-3}$\\
Decoder       & 4 Trans Block & 11.54M &$4.1\times 10^{-2}$\\
Duration Predictor & 2 Conv Layer & 0.40M & $4.2\times 10^{-4}$\\
Pitch Predictor & 5 Conv Layer & 1.64M & $4.8\times 10^{-3}$\\
Energy Predictor & 5 Conv Layer & 1.64M & $4.6\times 10^{-3}$\\
\bottomrule
\end{tabular}
\caption{Profiling of the model size and the inference latency of components in FastSpeech 2 model. RTF denotes the real-time factor which is the time (in seconds) required for the system to synthesize one second waveform. The RTF of each component is calculated by measuring the inference latency of the component and divided by the total generated audio length. The latency is measured with a single thread and a single core on an Intel Xeon CPU E5-2690 v4 @ 2.60 GHz, with 256 GB memory, and a batch size of 1.}
\label{tbl:analysis}
\end{table}

We have several observations from Table~\ref{tbl:analysis}: 1) The encoder and the decoder takes the most of the model size and the inference time. Therefore we mainly aim to reduce the encoder and the decoder size and perform architecture search to discover better and efficient architectures. 2) The predictors take about 1/3 of the total size and inference time. Therefore we will manually design the variance predictors with more lightweight operations rather than searching the architectures.

\subsection{Search Space Design}
\label{sec:search_space}
There are 4 feed-forward Transformer blocks~\cite{fastspeech} in both the encoder and the decoder in~\cite{fastspeech2}, where each feed-forward Transformer block contains a multi-head self-attention~(MHSA)~\cite{transformertts,fastspeech,fastspeech2} layer and a feed-forward network~(FFN)\footnote{FFN in~\cite{transformertts,fastspeech,fastspeech2} for TTS task consists of a 1D convolution layer and a fully connected layer.}. We use this encoder-decoder framework as our network backbone and set the number of layers in both the encoder and the decoder to $4$. For the variance predictors (duration, pitch and energy) in FastSpeech 2, our preliminary experiments show that removing the energy predictor makes very marginal performance drop in the voice quality. Accordingly, we directly remove the energy predictor in our design.

After setting the number of layers in each component, we search for different combinations of diverse architectures in the encoder and the decoder parts. We carefully design the candidate operations for our task: 1) LSTM is not considered due to the slow inference speed. 2) We decouple the original Transformer block to MHSA and FFN as separate operations. We adopt MHSA with different numbers of attention heads as $\{2, 4, 8\}$. 3) Considering that depthwise separable convolution~(SepConv)~\cite{sepconv} is much more memory and computation efficient compared to vanilla convolution, we use SepConv as an alternative of vanilla convolution. The parameter size of vanilla convolution is $K\times I_d\times O_d$ where $K$ is the kernel size, $I_d$ is the input dimension and $O_d$ is the output dimension. The size of SepConv is $K\times I_d + I_d\times O_d$. Following~\cite{seminas}, we adopt different kernel sizes of $\{1, 5, 9, 13, 17, 21, 25\}$. Finally, our candidate operations include $3+7+1=11$ different choices: MHSA with number of attention head in $\{2, 4, 8\}$, SepConv with kernel size in $\{1, 5, 9, 13, 17, 21, 25\}$ and FFN. This yields a search space of $11^{4+4}=11^8=214358881$ different candidate architectures.

To reduce the model size and latency of the variance predictors (the duration predictor and the pitch predictor), we directly replace the convolution operation in the variance predictors with SepConv in the same kernel size, without searching for other operations, which demonstrates to be very effective in our experiments.

\subsection{Search Algorithm}
\label{sec:search_algorithm} 
There have been many methods for searching neural architectures~\cite{nasnet,enas,nao,gbdtnas}. For our task, we adopt a very recent method~\cite{gbdtnas} which is based on accuracy prediction. It is efficient and effective, and well fits our task~(the chain-structure search space). Specifically, it uses a gradient boosting decision tree~(GBDT) trained on some architecture-accuracy pairs to predict the accuracy of other numerous candidate architectures. Then the architectures with top predicted accuracy are further evaluated by training on the training set and then evaluating on a held-out dev set. Finally, the architecture with the best evaluated accuracy is selected.

In our task, since the evaluation of the voice quality of a TTS system involves human labor, it is impractical to evaluate each candidate architecture during the search. We use the validation loss on the dev set as a proxy of the accuracy to guide the search\footnote{In non-autoregressive models (e.g., FastSpeech and FastSpeech 2), the valid loss on the dev set is highly correlated with the final quality, while in autogregressive  models it is not.}. Therefore, we search architectures with as small validation loss as possible.

\section{Experiments}
\label{sec:exp}

\subsection{Experimental Setup}
\textbf{Dataset} We evaluate our method on LJSpeech dataset~\cite{ljspeech}. LJSpeech contains $13100$ pairs of text and speech data with approximately $24$ hours of speech audio. We split the dataset into three parts: 12,900 samples as the training set, 100 samples as the dev set and 100 samples as the test set. Following~\cite{fastspeech2}, we convert the source text sequence into the phoneme sequence with an open-source grapheme-to-phoneme tool\footnote{https://github.com/Kyubyong/g2p}. We transform the raw waveform into mel-spectrograms following~\cite{fastspeech2}, and set the frame size and the hop size to 1024 and 256 with respect to the sample rate 22050.

\textbf{Search Configuration} For the GBDT-NAS algorithm, we follow the default setting and hyper-parameters in~\cite{gbdtnas}. The weight-sharing mechanism~\cite{oneshot,enas} adopted in~\cite{gbdtnas} trains and evaluates thousands of candidate architectures efficiently in a supernet which contains all the candidate architectures. Specifically, we train the supernet for 20k steps with a batch size of 28000 tokens per GPU and evaluate the validation loss of 1000 candidate architectures. Then, the GBDT predictor is trained on the 1000 architecture-loss pairs with 100 trees and 31 leaves per tree. Since the search space is moderate, we use the trained GBDT to predict the accuracy of all the candidate architectures in the search space, and re-evaluate the top 300 architectures by training them and evaluating their losses with the supernet. Finally, the architecture with the smallest validation loss is selected. The whole search process takes only 4 hours on 4 NVIDIA P40 GPU.

\textbf{Training and Inference} We train the searched TTS models on 4 NVIDIA P40 GPU, with bath size of 28000 tokens per GPU, for 100k steps. In the inference process, the output mel-spectrograms are transformed into audio samples using pre-trained Parallel WaveGAN~\cite{parawavegan} following~\cite{fastspeech2}.

\subsection{Results}
\begin{table}[htbp]
\small
\centering
\begin{tabular}{l|rr}
\toprule
Model            & \#Params     & CMOS \\
\midrule
FastSpeech 2     & 27.0M          & 0 \\
FastSpeech 2*    & 1.8M         & -0.230 \\
\midrule
LightSpeech      & 1.8M         & +0.04 \\
\bottomrule
\end{tabular}
\caption{The CMOS comparison between LightSpeech (our searched model), FastSpeech 2* (manually designed lightweight FastSpeech 2 model) and FastSpeech 2.}
\label{tbl:cmos}
\end{table}

\begin{table*}[htbp]
\centering
\small
\begin{tabular}{l|cc|rc|cc}
\toprule
Model          & \#Params & Compression Ratio & MACs & Ratio & Inference Speed (RTF) & Inference Speedup\\
\midrule
FastSpeech 2     & 27.0M    & /                 & 12.50G      & /       & $6.1\times 10^{-2}$   & /\\
\midrule
LightSpeech & \textbf{1.8M}   & \textbf{15x}  & \textbf{0.76G}  & \textbf{16x} & $9.3 \times 10^{-3}$ & \textbf{6.5x}\\
\bottomrule
\end{tabular}
\caption{The comparisons of model size, MACs and inference speed between LightSpeech and FastSpeech 2. The inference speed is measured in RTF (real time factor) with the same method as in Table~\ref{tbl:analysis}, using a single thread and a single core on an Intel Xeon CPU E5-2690 v4 @ 2.60 GHz. MACs is measured on a sample with input length $128$ and output length $740$.}
\label{tbl:speedup}
\end{table*}

\textbf{Audio Quality} To evaluate the quality of synthesized speech, we perform CMOS~\cite{cmos} evaluation on the test set. We compare our searched model (denoted as LightSpeech) with two baselines: 1) standard FastSpeech 2~\cite{fastspeech2}, and 2) manually designed lightweight FastSpeech 2 model (denoted as FastSpeech 2*), whose model size and inference latency can match to that of LightSpeech, with 2 feed-forward Transformer block in both the encoder and the decoder, hidden size 128 and filter size 256, no energy predictor, SepConv in the variance predictors. The other settings and configurations are the same as~\cite{fastspeech2}. The results are shown in Table~\ref{tbl:cmos}. We can see that the model discovered by our LightSpeech achieves comparable audio quality compared to FastSpeech 2 with no performance drop\footnote{CMOS within [-0.05, +0,05] is regarded as on par performance which is a common practice.}, while using much fewer number of parameters~(1.8M vs. 27M) which achieves 15x compression ratio. Meanwhile, manually designed lightweight FastSpeech 2 model~(FastSpeech2\*) has similar model size~(1.8M) but leads to severe performance drop compared to standard FastSpeech 2~(-0.230 CMOS). This demonstrates the advantages of our LightSpeech over human design in finding lightweight TTS model.

\textbf{Speedup and Complexity} Further, we measure the speedup and computation complexity in Table~\ref{tbl:speedup}. The inference speed on CPU in terms of RTF is reduced from $6.1\times 10^{-2}$ to $9.3\times 10^{-3}$ with 6.5x speedup\footnote{The RTF of FastSpeech 2 seems small on CPU ($6.1\times 10^{-2}$) since it is measured on a powerful server. However, on many devices with very constrained computation capability, the inference speed can be much slower. LightSpeech makes the deployment on these devices feasible with a 6.5x inference speedup.}. For computation complexity, we use the number of Multiply-Accumulate Operations~(MAC) to quantify the computation cost. LightSpeech has 16x fewer MACs than FastSpeech 2~(0.76G MACs vs. 12.50G MACs).

In summary, Table~\ref{tbl:cmos} and Table~\ref{tbl:speedup} show that the architecture discovered by LightSpeech achieves on par audio quality compared to FastSpeech 2 with 15x compression ratio, 16x fewer MACs and 6.5x inference speedup on CPU. Accordingly, it is more feasible to deploy in many resource constraint scenarios~(e.g., mobile platforms, embedded devices).

\subsection{Study and Analysis}
In this section, we study the effect of the designs proposed in Section~\ref{sec:search_space}.

\textbf{Shallowing} The manually designed FastSpeech 2* model shallows the FastSpeech model to 2 feed-forward Transformer block in both the encoder and the decoder. From the results in Table~\ref{tbl:cmos}, we see that the model size of FastSpeech 2* largely reduces from 27M to 1.8M while the corresponding audio quality severely drops with -0.230 CMOS compared to FastSpeech2. This indicates that simply compressing the model by shallowing the model results in performance drop.

\textbf{SepConv} We replace the convolution in the variance predictors with SepConv and show the loss of the duration predictor and the pitch predictor. The results are shown in Table~\ref{tbl:sepconv}. We can see that the loss does not drop, indicating the effectiveness of using SepConv to reduce model size while maintaining model capacity.
\begin{table}[htbp]
\centering
\small
\begin{tabular}{l|cc}
\toprule
Setting          & Duration Loss & Pitch Loss \\
\midrule
FastSpeech 2     & 0.12 & 0.85\\
FastSpeech 2 + SepConv & 0.12& 0.85\\
\bottomrule
\end{tabular}
\caption{Analysis on the effectiveness of replacing convolution with SepConv in the variance predictors.}
\label{tbl:sepconv}
\end{table}

\textbf{Search Space and NAS} To evaluate the effectiveness of our design of search space of the encoder and the decoder, we randomly sample 10 architectures from the search space and calculate their average validation losses. The results are in Table~\ref{tbl:searchspace}. We can see that random architectures achieve much lower loss compared to manually designed FastSpeech 2* (0.2753 vs.0.2956). This shows the effectiveness of our search space, which contains many promising architectures that are better than manually designed FastSpeech models. Further, architecture discovered by NAS (LightSpeech) achieves lower loss (0.2561), which is on par with that of FastSpeech 2 (0.2575), demonstrating the effectiveness of NAS.
\begin{table}[htbp]
\centering
\small
\begin{tabular}{l|cc}
\toprule
Setting          & \#Params & Mel Loss \\
\midrule
FastSpeech 2     & 27.0 M    & 0.2575\\
FastSpeech 2* & 1.8 M & 0.2956\\
\midrule
10 Random architectures & / & 0.2753 \\
LightSpeech & 1.8 M & 0.2561\\
\bottomrule
\end{tabular}
\caption{Analysis on the effectiveness of the search space and NAS.}
\label{tbl:searchspace}
\end{table}

\subsection{Discovered Architecture}
We show the final discovered architecture by our LightSpeech. The encoder consists of SepConv (k=5), SepConv (k=25), SepConv (k=13) and SepConv (k=9), and the decoder consists of SepConv (k=17), SepConv (k=21), SepConv (k=9), SepConv (k=13), where k is the kernel size. The hidden size is 256 as the same in~\cite{fastspeech2}. Other parts follow the descriptions in Section \ref{sec:search_space}.

\section{Conclusion}
In this paper, we propose LightSpeech, which leverages neural architecture search to discover lightweight and fast TTS models. We carefully analyze the memory and latency of each module in the original FastSpeech 2 model and then design corresponding improvements including the model backbone and the search space. Then we adopt GBDT-NAS to search well performing and efficient architectures. Experiments show that the discovered lightweight model achieves 15x compression ratio, 16x fewer MACs and 6.5x inference speedup on CPU with on par audio quality compared to FastSpeech 2. For future work, we will further combine NAS with other compression methods such as pruning and quantization for more efficient TTS models.

\bibliographystyle{IEEEbib}
\bibliography{my}

\end{document}